# Best insertion algorithm for resource-constrained project scheduling problem

*Igor Pesek, Janez Žerovnik*
*Institute of mathematics, physics and mechanics*
*Jadranska cesta 19, Ljubljana, Slovenia*
*e-mails: igor.pesek@imfm.uni-lj.si*
*janez.zerovnik@imfm.uni-lj.si*

**Abstract**: This paper considers heuristics for well known resource-constrained project scheduling problem (RCPSP). First a feasible schedule is constructed using randomized best insertion algorithm. The construction is followed by a local search where a new solution is generated as follows: first we randomly delete *m* activities from the list, which are then reinserted in the list in consecutive order. At the end of run, the schedule with the minimum makespan is selected. Experimental work shows very good results on standard test instances found in PSPLIB.

## 1. Introduction

The resource-constrained project scheduling problem (RCPSP) can be stated as follows. Given are *n* activities *1, ... , n* and r renewable resources. A constant amount of $R_k$ units of resource *k* is available at any time. Activity *i* must be processed for $p_i$ time units; preemption is not allowed. During this time period a constant amount of $r_{i,k}$ units of resource *k* is occupied. The values $R_k$, $P_i$ and $r_{ik}$ are supposed to be non-negative integers. Furthermore, there are precedence relations defined between activities. The objective is to determine starting times $s_i$ for the activities *i = 1, ... , n* in such a way that

- at each time *t* the total resource demand is less than or equal to the resource availability for each resource type,
- the given precedence constraints are fulfilled and,
- the makespan $C_{\max} = \max_{i=1}^{n} c_i$, where $c_i = s_i + p_i$, is minimized.

As a generalization of the job-shop scheduling problem the RCPSP is *NP*-hard in the strong sense.

In recent years, many works have been published including the main research on RCPSP (Kolisch and Hartmann [1], Özdamar and Ulusoy [2]) and great advances have been made in the solving procedures which take into account two different approaches: optimal and heuristic. The optimal approach includes methods such as dynamic programming (Carruthers and Battersby [3]) and implicit enumeration with branch and bound (Talbot and Patterson [4]). Nevertheless, the NP-hard nature of the problem makes it difficult to solve realistic sized projects, in such a way that, in practice, the use of heuristics is necessary. Therefore, besides exact algorithms many authors have developed heuristics for the RCPSP as the only feasible method of handling practical resource constrained project scheduling problems (for survey see [5]).

In this paper we first present a procedure for generating a feasible initial schedule using an adaptation of the randomized best insertion algorithm [6]. The construction is followed by the local search, in which we first delete $m$ activities and then reinsert them back in to the schedule.

## 2. The heuristic

### 2.1 Building initial solution

Majority of methods start with none of the jobs being scheduled. Subsequently, a single schedule is constructed by selecting a subset of jobs in each step and assigning starting times to these jobs until all jobs have been considered. This process is controlled by the scheduling scheme as well as priority rules with the latter being used for ranking the jobs.

We approach generating initial schedule differently. In the first step, we randomly choose $m$ activities from all the possible activities regardless to their constraints and optimally solve this partial schedule by a branch-and-bound algorithm [7]. The number $m$ is arbitrarily chosen number enabling us to optimally solve this subset very fast; usually we set $m$ between 10 and 25.

In the next step remaining activities were randomly chosen and reinserted in the schedule. The position of the insertion was obtained by searching all the positions in the schedule. Between the positions, which are feasible and make the minimal makespan change, one is randomly chosen.

Because some activities are inserted in to the schedule before all their predecessors are, we must check only the activities which are already in the schedule. Therefore we must consider also transitive predecessors preventing the schedule to become infeasible.

Similarly we must also check that at some position there are also all the direct and transitive successors behind reinserted activity.

## 2.2    Local search

Our local search was inspired by the good results that were achieved with a similar approach for TSP problem [6]. The main idea is to remove fixed number of vertices from the schedule and insert them back in to the schedule. Whole procedure is very similar to the method described in previous section, with one important distinction. Here we already have a schedule, whereas in the previous method we are building one.

Let us now describe the algorithm. We presume that we already have initial schedule and it is stored in *S*. Each iteration of our local search consists of the following steps.

```
1. S' = S
2. remove m activities from S'
3. for each removed activity
   3.1.   check every position in partial schedule
   3.2    choose the position that makes the minimal makespan (if there
          are more such positions take one randomly)
   3.3    insert activity to that position
4. Compare schedule S and S', choose the schedule with smaller total
   makespan and store it in the S
```

If the total makespan of schedules *S* and *S'* in step 4 is same, then we keep the original schedule. After the fixed number iterations with no improvement on the total makespan, we discard the original schedule and use instead the new schedule in next iterations. This way the probability of being caught in local minimum is smaller.

Interesting question is how many activities should be removed and then reinserted back to the schedule. Our experiments show that the best results are obtained when the number is around 1/10 of activities of project. Removing too few activities does not have the desired effect, since we are not taking the advantages of this approach. On other hand, the number above proposed number usually disrupts the original schedule in such a way that no good results are obtained within comparable times.

One of the advantages of this method is its adaptiveness to systems or projects, where activities are removed or added dynamically. For example in factories situations can occur which need some modification and need for dynamic change of process is needed frequently. In such situation, other heuristics and meta-heuristics, for example tabu search or simulated annealing, must start from beginning. They need to build initial solution again and begin searching for the best one with that schedule, therefore forgetting all the previous work on that project. For our method, a small change in the instance does not causes a considerable change, as it just inserts added activity like any other activity which is removed and reinserted. Therefore it is very appropriate for dynamic systems in general.

## 2.3 Combining with metaheuristics

Our method can also be very successfully combined with all known metaheuristic paradigms. Since we randomly choose activities that we remove and later reinsert, the definition of neighborhood has to be chosen accordingly. In every iteration we select $k$ of $m$-activities to be removed and reinserted.

## 3. Computational results

In this section we present some computational results for the local search methods developed in the previous section. We investigated the performance of our algorithm with respect to various benchmark instances of project Scheduling Library (PSPLIB) [8].

The atomic move for all three metaheuristic paradigms is the same. Neighborhood is defined on all pairs of activities $i$ and $j$, which can swap their position and all the constraints remain satisfied. We limited all searches to 3000 and 5000 iterations for last instance, respectively. Additional parameters were set: for tabu search we set minimum tenure to 10 and maximum tenure to 15; for simulated annealing we set starting temperature to 2.0 and cooling ratio to 0.99 with 100 samples sampled at each temperature. We did not spend a lot of time to tune the parameters of the algorithms, because this was not the focus of our research. Note also, that our implementation of hill climbing method chooses solution randomly at each iteration between all the improving solutions in the neighborhood. All the metaheuristics and also our methods were implemented and tested in EasyLocal++ framework, which provides environment for testing such methods [9].

All runs were limited to 3000 iterations with the instance j9010_5.sm, which consists of 90 jobs and 4 available resources. Computed lower bound for this instance is 78. In the third column time needed for completing all the iterations is stated. Note, that this is not the time, when best solution has been found. In fact many heuristics has found best solution very quickly, e.g. multi move tabu search found his best solution already after 5 iteration.

**Table 1: Instance j9010_5.sm, 3000 iterations, known best solution = 78**

| Heuristics | Best solution | t(s) |
|---|---|---|
| Tabu search – MultiMove (MM) | 78 | 7082 |
| Tabu search – Remove and reinsert (RAR) | 78 | 19082 |
| Simulated Annealing - MM | 78 | 1,5 |
| Simulated Annealing - RAR | 78 | 125 |
| Hill Climbing – MM | 78 | 1 |
| Hill Climbing – RAR | 78 | 120 |
| Remove and Reinsert with 5 activities | 78 | 52 |
| Remove and Reinsert with 10 activities | 78 | 100 |

With the instance j9021_6.sm all runs were limited to 3000 iterations. The problem consists of 90 jobs and 4 available resources. Computed lower bound for this instance is 95, but has not been verified. In fact best found solution = 106 so far is due to D. Debels and M. Vanhoucke and has been found in 2005.

**Table 2: Instance j9021_6.sm, 3000 iterations, known best solution = 106**

| Heuristics | Best solution | t(s) |
|---|---|---|
| Tabu search – MultiMove (MM) | 117 | 6465 |
| Tabu search – Remove and reinsert (RAR) | 111 | 18098 |
| Simulated Annealing - MM | 121 | 2 |
| Simulated Annealing - RAR | 116 | 122 |
| Hill Climbing – MM | 117 | 2 |
| Hill Climbing – RAR | 111 | 171 |
| Remove and Reinsert with 5 activities | 118 | 62 |
| Remove and Reinsert with 10 activities | 114 | 97 |

Next we computed the solutions for j1201_2.sm, where problem consists of 120 jobs and 4 available resources. Best solution found, value is 109, is due to P. Laborie in 2005. Since the problem is larger we increased the number of iterations to 5000.

**Table 3: Instance j1201_2.sm, 5000 iterations, best known solution = 109**

| Heuristics | Best solution | t(s) |
|---|---|---|
| Tabu search – MultiMove (MM) | 113 | 26899 |
| Tabu search – Remove and reinsert (RAR) | 111 | 72373 |
| Simulated Annealing - MM | 123 | 3 |
| Simulated Annealing - RAR | 118 | 353 |
| Hill Climbing – MM | 118 | 3 |
| Hill Climbing – RAR | 116 | 320 |
| Remove and Reinsert with 5 activities | 114 | 154 |
| Remove and Reinsert with 10 activities | 112 | 297 |

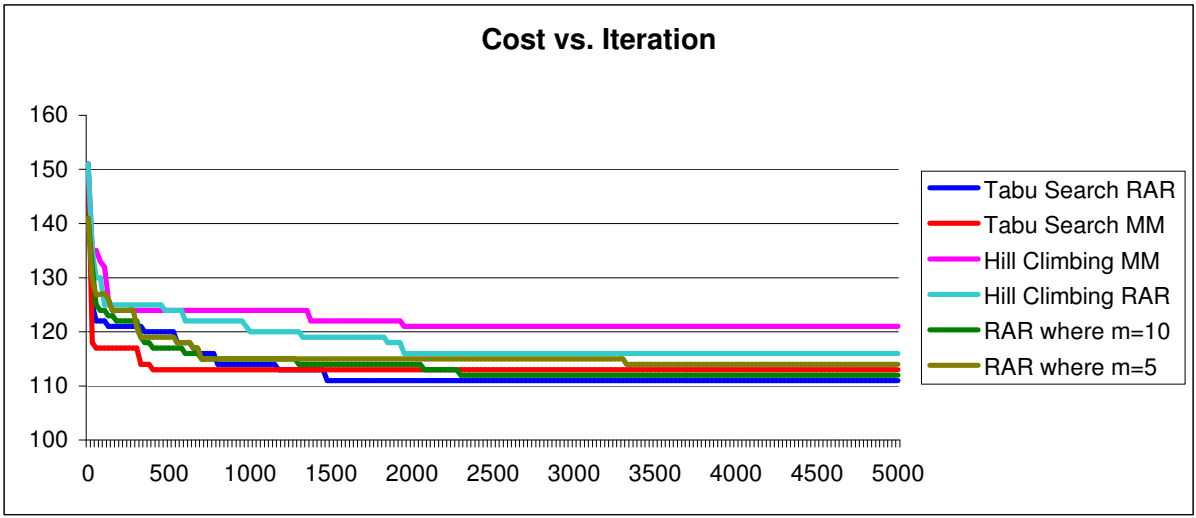

**Figure 1 Cost vs. iteration for j1201_2.sm instance**

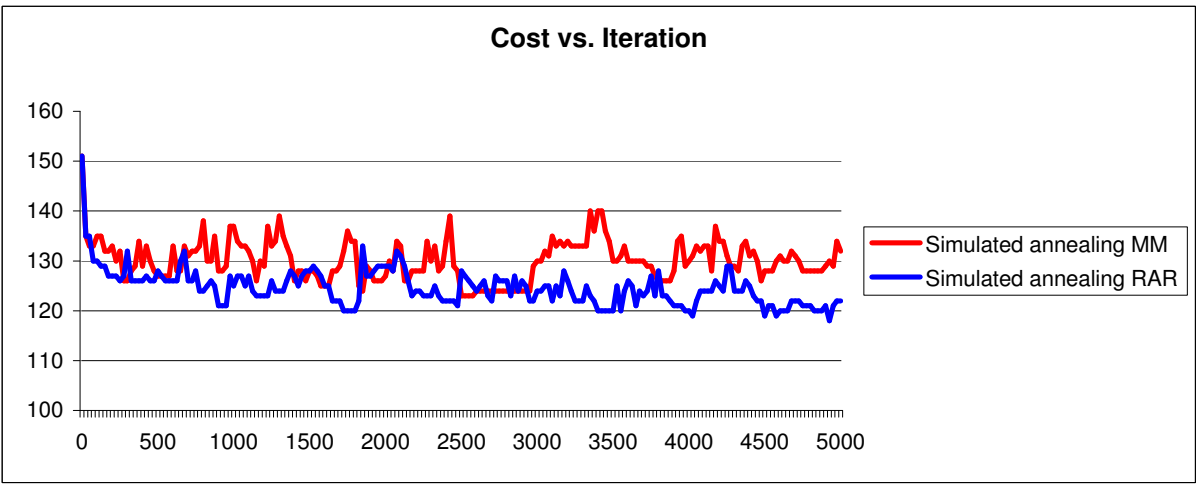

**Figure 2 Cost vs. iteration for j1201_2.sm instance**

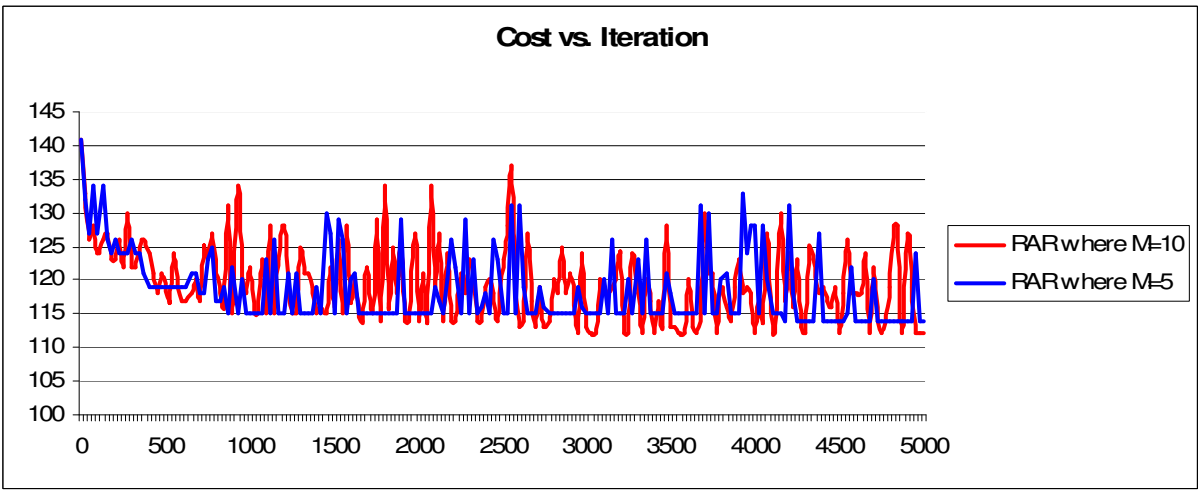

**Figure 3 Cost vs. iteration for j1201_2.sm instance**

In figures 1, 2 and 3 the graph for the instance j1201_2.sm shows how the quality of the solution is improving with the number of iterations increasing.

Experiments show, that our method performs very well and is competitive to other similar methods. As already stated in previous section our method yields very good results, when ~1/10 of all activities are removed and later reinserted.

## 4. Conclusion

In this paper we introduced a randomized best insertion algorithm, which performs very well in both static and dynamic project scheduling problems. Our code and approach was tested on some of instances from PSPLIB library. For these problems, we succeeded to find practically usable schedules. The preliminary results of the computational experiments suggest that the proposed algorithm is a very competitive heuristic and yields better results than several heuristics presented in the literature.

At last, the algorithm is easily adaptable to optimize other objective functions based on the project duration and most of all it is suitable to solve real dynamic problems.